\documentclass[showpacs,superscriptaddress,amsmath,amssymb,twocolumn]{revtex4}
\usepackage{graphicx}
\usepackage{bm}

\begin{document}

\title{Dynamics of matter-wave solutions of Bose-Einstein
 condensates in a homogeneous gravitational field}

\author{Etienne Wamba}
\email{wambaetienne@yahoo.fr}

\affiliation{Laboratory of Mechanics, Department of Physics, Faculty
of Science, University of Yaound\'{e} I, P.O. Box 812, Yaound\'{e},
Cameroon}

\affiliation{The Abdus Salam International Centre for Theoretical
Physics, P.O. Box 586, Strada Costiera 11, I-34014, Trieste, Italy}

\affiliation{Department of Physics, School of Physical, Chemical and
Applied Sciences, Pondicherry University, Pudhucherry 605014, India}

\author{Hermano Velten}
\email{velten@physik.uni-bielefeld.de}


\affiliation{Fakult\"{a}t f\"{u}r Physik, Universit\"{a}t Bielefeld,
Postfach 100131, 33501 Bielefeld, Germany}

\author{Alidou Mohamadou}
\email{mohdoufr@yahoo.fr}

\affiliation{The Abdus Salam International Centre for Theoretical
Physics, P.O. Box 586, Strada Costiera 11, I-34014, Trieste, Italy}

\affiliation{ Condensed Matter Laboratory, Department of Physics,
Faculty of Science, University of Douala, P.O. Box 24157, Douala,
Cameroon}

\author{Timol\'{e}on C. Kofan\'{e}}
\email{tckofane@yahoo.com}

\affiliation{Laboratory of Mechanics, Department of Physics, Faculty
of Science, University of Yaound\'{e} I, P.O. Box 812, Yaound\'{e},
Cameroon}

\author{K. Porsezian}
\email{ponz.phy@pondiuni.edu.in}

\affiliation{Department of Physics, School of Physical, Chemical and
Applied Sciences, Pondicherry University, Pudhucherry 605014, India}

\date{\today}

\begin{abstract}
We find a matter-wave solution of Bose-Einstein condensates trapped
in a harmonic-oscillator potential and subjected to a homogeneous
gravitational field, by means of the extended tanh-function method.
This solution has as special cases the bright and dark solitons. We
investigate the dynamics and the kinematics of these solutions, and
the role of gravity is sketched. It is shown that these solutions
can be generated and manipulated by controlling the \emph{s}-wave
scattering length, without changing the strengths of the magnetic
and gravitational fields.
\\

\textit{\text{Keywords}:\ Extended tanh-function method,
Gross-Pitaevskii equation, gravitational field, soliton
solutions.}
\end{abstract}


\maketitle

\section{Introduction}

Bose-Einstein condensation is a very general physical phenomenon
which takes place in the systems of bosonic atoms at ultralow
temperatures, as well as in optical wave systems \cite{zakharov}. It
appears in the fields of condensed matter, atomic and elementary
particle physics and astrophysics \cite{griffin}. The majority of
important features of the condensation in such diverse systems can
be captured by the Gross-Pitaevskii (GP) equation, which is a
variant of the nonlinear Schr\"{o}dinger equation with a trap
potential. Due to the nonlinearities arising from the interatomic
interactions and due to the presence of a confining potential in the
GP equation, many studies have been performed either by solving the
corresponding GP equation numerically or by using perturbative
methods (see \cite{wamba6,wamba8} and references therein). However
the construction of exact solutions of the GP equation can help to
advance our understanding of the various physical phenomena governed
by this nonlinear equation. For instance, the exact analytical
solutions can contribute to select the experimental parameters, to
analyze the stability of Bose-Einstein condensates (BECs) and to
check numerical analysis of this nonlinear equation \cite{zhao}.
Therefore, the construction of the exact solution of the GP equation
is one of the most relevant challenges to take up. To this end, many
works have proposed several methods to exactly solve the GP
equation, such as the Darboux transformation method, the
hyperbolic-function method, the elliptic-functions method, the
inverse-scattering method, the generalized $(G'/G)$-expansion
method, the Hirota bilinear method, the Painlev\'{e} analysis, the
Lie symmetry, the tanh-function method and the extended
tanh-function method (see for instance \cite{methods} and references
therein). Finding the nontrivial seed solutions that mimics the
sought solutions of the GP equation always remains a hard task.

Recently, there have been some experimental reports on solitons in
the quasi one-dimensional (1D) system which have been observed by
magnetically tuning the interparticle interaction from repulsive to
attractive. Experimentally, for the realization of 1D systems, one
should make the radial frequency much larger than the axial
frequency and strongly confine the radial motion. The state in this
cylindrical harmonic-oscillator trap represents a nonlinear matter
wave.

The formation and propagation of matter waves, such as dark
solitons, bright solitons, four-wave mixing and moving or stationary
gap solitons \cite{murali} are among the most interesting dynamical
features in Bose-Einstein condensation. Of course, such dynamics
depends on the types of interactions in which the system is
subjected to. For atoms in the nK-mK temperature regime, the effect
of the Earth's gravitational field is by no means negligible
especially in the case of magnetic trapping. The gravitational field
plays a subtle role in the topological formation of stable vortices
within the field reverse time in BEC experiments with heavier atoms
like $^{87}$Rb \cite{kawaguchi}. It has been shown that the
gravitational field can change the propagation trail of the bright
soliton trains without changing the peak and width of the soliton in
the parabolic background \cite{zhao,wamba3}. Moreover, the presence
of a homogeneous gravitational field can decrease the condensation
temperature of Bose gases \cite{rivas}.

\par
BECs also appear in the astrophysical context. For example, one has
recently argued that dark matter \cite{bertone}, an unknown
component which corresponds to approximately 25 \% of the energy
content of the Universe, is currently under the form of a BEC
\cite{velten-sin1994-harko2008}. In this approach, bosonic dark
matter particles underwent a phase transition forming a BEC at some
point during the evolution of the Universe. As a consequence, one
opens the possibility of admitting the existence of BECs subjected
to many gravitational effects.  This seems to be a very promising
line of research for the next years.
\par
In this paper, we study the dynamics of BECs in a magnetic trap in
the presence of gravity. We construct an exact solution of the GP
equation which has as special cases the well-known bright and dark
solitons. The work is organized as follows. In Sec. \ref{sec2}, we
present the model. Then in Sec. \ref{sec3}, by utilizing the
extended tanh-function method, we find the various soliton
solutions. In Sec. \ref{sec4} we examine the dynamical properties of
these solutions.
Finally in Sec. \ref{sec5}, we summarize our results and conclude the work.\\

\section{The model}\label{sec2}

As is well known, in the mean field approximation, the full dynamics
of a BEC in a trap potential $V (\mathbf{r})$ satisfies the
time-dependent Gross-Pitaevskii (GP) equation \cite{trombettoni,
dalfovo, abdullaev}
\begin{equation}
\mathrm{i}\hbar \Psi_{t}=-\frac{\hbar^{2}}{2 m}\nabla^{2}\Psi+ V
(\mathbf{r})\Psi + g_{0} |\Psi|^{2} \Psi, \label{eq00}
\end{equation}
where $g_{0}=\frac{4\pi \hbar a_{s}}{m}$, with $m$ being the atomic
mass and $a_{s}$ the \emph{s}-wave scattering length which can be
either \emph{positive} (the case of $^{87}$Rb atoms with
$a_{s}=5.45\pm 0.26$ nm and then repulsive
interactions)\cite{julienne} or \emph{negative} (the case of
$^{7}$Li atoms with $a_{s}=-1.45$ nm and then attractive
interactions) \cite{abdullaev}. It has been shown that, for alkali
atoms at least, the effect of gravity is non-negligible
\cite{camacho,legget,ensher}. We recall that in the usual
experimental traps, the atomic clouds are confined with the help of
laser or magnetic trapping. For alkali atoms such as rubidium, which
is the most used boson for experiments on cold atoms, most of the
existing traps can be well approximated by a three-dimensional
harmonic oscillator \cite{dalfovo}. We consider a trapped Bose gas
of non-relativistic particles immersed in a Newtonian gravitational
field. In this case, the external confining potential must take into
account the contribution of both the magnetic trapping field and the
gravitational field. It is given by the sum of the quadratic and
gravitational potentials generated by these respective fields
\cite{rivas,camacho,kulikov}:
\begin{equation}
V(\mathbf{r})=\frac{m}{2}\big(\omega_{x}^{2} x^{2} + \omega_{y}^{2}
y^{2} + \omega_{z}^{2} z^{2} \big) + m g z, \label{eq00a}
\end{equation}
where $\omega_{l}$, with $l=x,y,z$, denote the frequencies of the
harmonic oscillator along the coordinate axes. The parameter $g$,
which represents the acceleration of gravity, is taken as constant
since the gravitational field is homogeneous. The gravitational
potential represented by the last term of Eq. (\ref{eq00a}) is also
present in the case of the tilted trap
\cite{trombettoni,anderson-zhangw}.

In general, the harmonic-oscillator potential represented by the
first term in Eq. (\ref{eq00a}) is anisotropic, i.e., the trapping
frequencies $\omega_{l}$ are all different.
For a trap that is axially symmetric along the $z$- axis, we have
$\omega_{x}=\omega_{y}\equiv \omega_{\bot}$. In such a case,
$\omega_{z}$ is referred to as the longitudinal frequency (along the
axial direction) while $\omega_{\bot}$ is the radial frequency of
the anisotropic harmonic trap. Then Eq. (\ref{eq00}) reduces to the
following three-dimensional GP equation:
\begin{equation}
\mathrm{i}\hbar \Psi_{t}=-\frac{\hbar^{2}}{2 m}\nabla^{2}\Psi+
[\frac{1}{2}m (\omega_{\bot}^{2} r_{\bot}^{2}+\omega_{z}^{2} z^{2})
+ m g z] \Psi + g_{0} |\Psi|^{2} \Psi, \label{eq02}
\end{equation}
where $r_{\bot}=\sqrt{x^{2}+y^{2}}$ is the radial distance. The
radial motion can be strongly confined by making the radial
frequency much larger than the axial frequency, i.e. $\omega_{\bot}
\gg \omega_{z}$. In this case the condensate is cigar-shaped, and
owing to that \cite{zhang-abdu}, one can take
$\Psi(\mathbf{r},t)=\phi_{0}(r_{\bot})\psi(z,t)$, where
$\phi_{0}=\sqrt{\frac{1}{\pi
a_{\bot}^{2}}}\exp(-\frac{r_{\bot}^{2}}{2a_{\bot}^{2}})$ is the
ground state of the radial problem, with
$a_{\bot}=\sqrt{\hbar/m\omega_{\bot}}$. Then multiplying both sides
of the GP equation (\ref{eq02}) by $\phi_{0}^{*}$ and integrating
over the transverse variable $r_{\bot}$ we obtain a
quasi-one-dimensional GP equation in the form:
\begin{equation}
\mathrm{i}\hbar \psi_{t}=-\frac{\hbar^{2}}{2 m}\psi_{zz}+
\big(\frac{1}{2}m \omega_{z}^{2} z^{2} + m g z\big) \psi + \upsilon
|\psi|^{2} \psi. \label{eq01}
\end{equation}
Thus, the GP equation (\ref{eq01}) describes the dynamics of trapped
quasi-1D cigar-shaped Bose gases at the mean-field level.
In this equation, the strength of the atom-atom interaction becomes
$\upsilon=2\hbar\omega_{\bot} a_{s}$. The $s$-wave scattering length
$a_{s}$ can be managed through the Feshbach resonance technique
\cite{feshbach}.
Additionally, the effect of the acceleration of gravity $g$ in the
BEC experiment can be tiny varied or even cancelled in drop tower
experiments \cite{nandi}. In the  astrophysical context, we know
that BEC might exist (in a speculative scenario) and can be
subjected to very high gravitational fields (close to black holes,
for example). This suggests the possibility to vary the
gravitational field.
Hence one has some freedom in choosing the physical parameters of
the system. In what follows, we set $\frac{\hbar}{2m}=c$,
$\frac{1}{2}m \omega_{z}^{2}=\hbar\alpha$, $m g=\hbar\lambda$, and
$\upsilon=\hbar\nu(t)$.

\smallskip
We follow the ideas of Refs. \cite{wamba6,wamba8} and seek the exact
solitonic solutions of Eq. (\ref{eq01}) within the extended
tanh-function method. We therefore introduce the auxiliary equation
\begin{equation}
u'^{\,2}=c_{0}+c_{1}u+c_{2}u^{2}+c_{3}u^{3}+c_{4}u^{4},
\label{eq24}
\end{equation}
where $u'=\frac{\mathrm{d} u}{\mathrm{d}\zeta}$, $\zeta=\zeta(z,t)$
and $c_{0}$, $c_{1}$, $c_{2}$, $c_{3}$, $c_{4}$ are some real
constants.
Let $a$ be an arbitrary real constant. One may verify that Eq.
(\ref{eq24}) admits for $c_{0}=0, \: c_{1}=0, \: c_{2}>0$, and
$c_{3}, c_{4}=arbitrary$, the following solution:
\begin{equation}
u(\zeta)=\frac{4 c_{2}}{-2 c_{3}+(c_{3}^{2}-4c_{2}c_{4})e^{\delta\sqrt{c_{2}}\,(\zeta-a)}+e^{-\delta\sqrt{c_{2}}\,(\zeta-a)}}, \label{eq25}\end{equation}\\
where $\delta =\pm 1$. It is worth noting that this solution
presents some singularities for $c_{3}^{2}-4c_{2}c_{4}<0$.

We let $\zeta(z,t)=p(t) z+q(t)$ and use the transformation
\begin{equation}
\psi(z,t)=[f(t)+h(t)u(\zeta)]e^{\mathrm{i}[\beta(t) z^{2} + k(t)
z+\Omega(t)]}. \label{eq28}
\end{equation}
The parameter $\alpha$ is called the steepness of the harmonic
trapping potential, in reference to the potential energy of a
perfectly elastic spring. $\beta(t)$ is twice the chirp and
$\Omega(t)$ is the linear phase.
Next, we substitute Eqs. (\ref{eq24}) and (\ref{eq28}) into Eq.
(\ref{eq01}), and set the real and imaginary parts of the resulting
equation to zero. Then collecting coefficients of powers of
$z^{n}u^{n'}$ ($n=0,1; \: n'=0,1,2,3,4,5$) and setting each of them
to zero, we obtain the following set of over-determined ordinary
differential equations:
\begin{equation}
\begin{split}
&f'+ 2c \beta f=0,\: h'+ 2c \beta h=0,\\
&h (p'+ 4c \beta p)=0,\: h (q'+2c k p)=0,\\
&h (k'+ 4c \beta k+\lambda)=0,\: f (k'+ 4c \beta k+\lambda)=0,\\
&f (\beta'+4c\,\beta^{2}+\alpha)=0,\: h (\beta'+4c\,\beta^{2}+\alpha)=0,\\
&h (2c\, p^{2}c_{4}-\nu h^{2})=0,\: \frac{3}{2}h(cp^{2}c_{3}-2 \nu f h)=0,\\
&2f (\Omega'+c\,k^{2}+\nu f^{2})-c\,h p^{2}c_{1}=0,\\
&h(\Omega'+c\,k^{2}+3\nu f^{2}-c\, p^{2}c_{2})=0. \label{eq29o}
\end{split}
\end{equation}
In this set of equations, the prime stands for the derivative with
respect to the time $t$. Since $c_{1}=0$ is the constraint on the
solution (\ref{eq25}), the eleventh equation of the set of equations
(\ref{eq29o}) may read $f (\Omega'+c\,k^{2}+\nu f^{2})=0$. Applying
different constraints between the parameters in Eq. (\ref{eq25}),
the above equations may yield to various solutions. Thus to solve
Eq. (\ref{eq29o}), we consider two situations that have interesting
implications on the matter wave solution, namely when $c_{3}=0$ and
$c_{3}\neq 0$.

\section{The solutions} \label{sec3}

\subsection{The bright solitons}

For $c_{3}=0$, we get $f(t)=0$, and then we may reduce the above set
of equations (\ref{eq29o}) to
\begin{equation}
\begin{split}
&h'+ 2c \beta h=0,\: p'+ 4c \beta p=0,\: q'+2c k p=0,\\
&k'+ 4c \beta k+\lambda=0,\: \beta'+4c\beta^{2}+\alpha=0,\\
&\Omega'+ck^{2}-c p^{2}c_{2}=0,\: 2cp^{2}c_{4}-\nu h^{2}=0.
\label{eq29}
\end{split}
\end{equation}
In the case of time-independent magnetic and gravitational fields,
say $\alpha$ and $\lambda$ are constants, we may choose the
parameter $\beta(t)$ in such a way that the above set is easily
solvable. We consider the following three examples.

(1) When $\beta(t)$ is constant, solving the above set of
equations yields:
\begin{equation}
\begin{split}
&\beta=\sqrt{-\frac{\alpha}{4c}},\, p(t)=p_{0} \ell(t)^{-1},\, h(t)=h_{0} \ell(t)^{-1/2},\\
&k(t)=-\frac{\lambda}{4c \beta}+k_{1} \ell(t)^{-1},\, \nu(t)=\nu_{0} \ell(t)^{-1},\, \ell(t)= e^{-4c \beta t},\\
&q(t)=\left(-\frac{\lambda}{8c\beta^{2}}+\frac{k_{1}}{4\beta}e^{-4c
\beta t}\right) p(t)+q_{1},\\
&\Omega(t)=-\frac{\lambda^{2}}{16c
\beta^{2}}t-\frac{c_{2}p_{0}^{2}-k_{1}^{2}}{8\beta}e^{-8c \beta
t}-\frac{\lambda k_{1}}{8c\beta^{2}}\ell(t)^{-1}+\Omega_{1}.
\label{eq30}
\end{split}
\end{equation}
In this set of equations, $\nu_{0}$ and $h_{0}=\pm(2c
\frac{c_{4}}{\nu_{0}} p_{0}^{2})^{1/2}$ are the initial values of
$\nu(t)$ and $h(t)$, respectively. The parameter $p_{0}$ is a
nonzero real constant while $k_{1}$, $q_{1}$ and $\Omega_{1}$ are
some arbitrary real constants. These results implicitly assume
$\nu_{0} c_{4}>0$ and particularly $\alpha<0$, in view of the above
expressions of $\beta$ and $h_{0}$. Thus the solution in this case
corresponds to expulsive magnetic trapping potential. Furthermore,
it demands $c_{4}\neq 0$ in order for the amplitude coefficient $h$
to be nonzero.

(2) When
$\beta(t)=-\sqrt{\frac{\alpha}{4c}}\,\tan(2\sqrt{c\alpha}\,t)$,
solving the set of equations (\ref{eq29}), we come to the
following results
\begin{equation}
\begin{split}
p(t)=& p_{0} \ell(t)^{-1},\,
h(t)= h_{0} \ell(t)^{-1/2},\\
k(t)=& -\frac{\lambda}{2\sqrt{c\alpha}}\,\tan(2\sqrt{c\alpha}\,t)+k_{1} \ell(t)^{-1},\\
q(t)=& \left[-k_{1}\sqrt{\frac{c}{\alpha}}\sin(2\sqrt{c\alpha}\,t)+\frac{\lambda}{2\alpha}\right] p(t)+q_{1},\\
\Omega(t)=& \frac{\lambda^{2}}{4\alpha}t-\frac{\lambda^{2}-4c\alpha (c_{2} p_{0}^{2}-k_{1}^{2})}{8\alpha\sqrt{c\alpha}}\,\tan(2\sqrt{c\alpha}\,t)\\
&+\frac{\lambda k_{1}}{2\alpha} \ell(t)^{-1}+\Omega_{1},\\
\nu(t)=& \nu_{0} \ell(t)^{-1},\, \ell(t)=
|\cos(2\sqrt{c\alpha}\,t)|. \label{eq31}
\end{split}
\end{equation}
These results also assume $\nu_{0} c_{4}>0$, with $c_{4}\neq 0$ and
especially $\alpha>0$. Hence the solution in this second case
corresponds to attractive magnetic trapping potential. It signals
singularities at any $t=\frac{(2n+1)\pi}{4\sqrt{c\alpha}}$, with $n$
being any positive integer. In such case we consider only safe times
defined by $t<\frac{\pi}{4\sqrt{c\alpha}}$ to prevent singularities.
This assumption does not alter the validity of our approach since
the lifetime of a condensate is small in general. Moreover most of
experiments are done with weak magnetic field, i.e., with small
values of the parameter $|\alpha|$. In this situation, the cut-off
time $t_{1}=\frac{\pi}{4\sqrt{c\alpha}}$ is long enough to allow the
observation of the matter-wave propagation in the condensate.

(3) When
$\beta(t)=\sqrt{-\frac{\alpha}{4c}}\,\tanh(2\sqrt{-c\alpha}\,t)$,
solving the set of equations (\ref{eq29}), we obtain the following
results
\begin{equation}
\begin{split}
p(t)=& p_{0} \ell(t)^{-1},\,
h(t)= h_{0} \ell(t)^{-1/2},\\
k(t)=& -\frac{\lambda}{2\sqrt{-c\alpha}}\,\tanh(2\sqrt{-c\alpha}\,t)+k_{1} \ell(t)^{-1},\\
q(t)=& \left(k_{1}\sqrt{-\frac{c}{\alpha}}\,e^{-2\sqrt{-c\alpha}\,t}+\frac{\lambda}{2\alpha}\right) p(t)+q_{1},\\
\Omega(t)=& \frac{\lambda^{2}}{4\alpha}t-\frac{\lambda^{2}-4c\alpha (c_{2} p_{0}^{2}-k_{1}^{2})}{8\alpha\sqrt{-c\alpha}}\,\tanh(2\sqrt{-c\alpha}\,t)\\
&+\frac{\lambda k_{1}}{2\alpha} \ell(t)^{-1}+\Omega_{1},\\
\nu(t)=& \nu_{0} \ell(t)^{-1},\, \ell(t)=
\cosh(2\sqrt{-c\alpha}\,t). \label{eq32}
\end{split}
\end{equation}
Likewise these results also assume $\nu_{0} c_{4}>0$, with
$c_{4}\neq 0$ but especially $\alpha<0$. Hence the solution in this
third case corresponds to expulsive magnetic trapping potential.

\begin{figure}[tbh]
\centering
\includegraphics[width=2.0in]{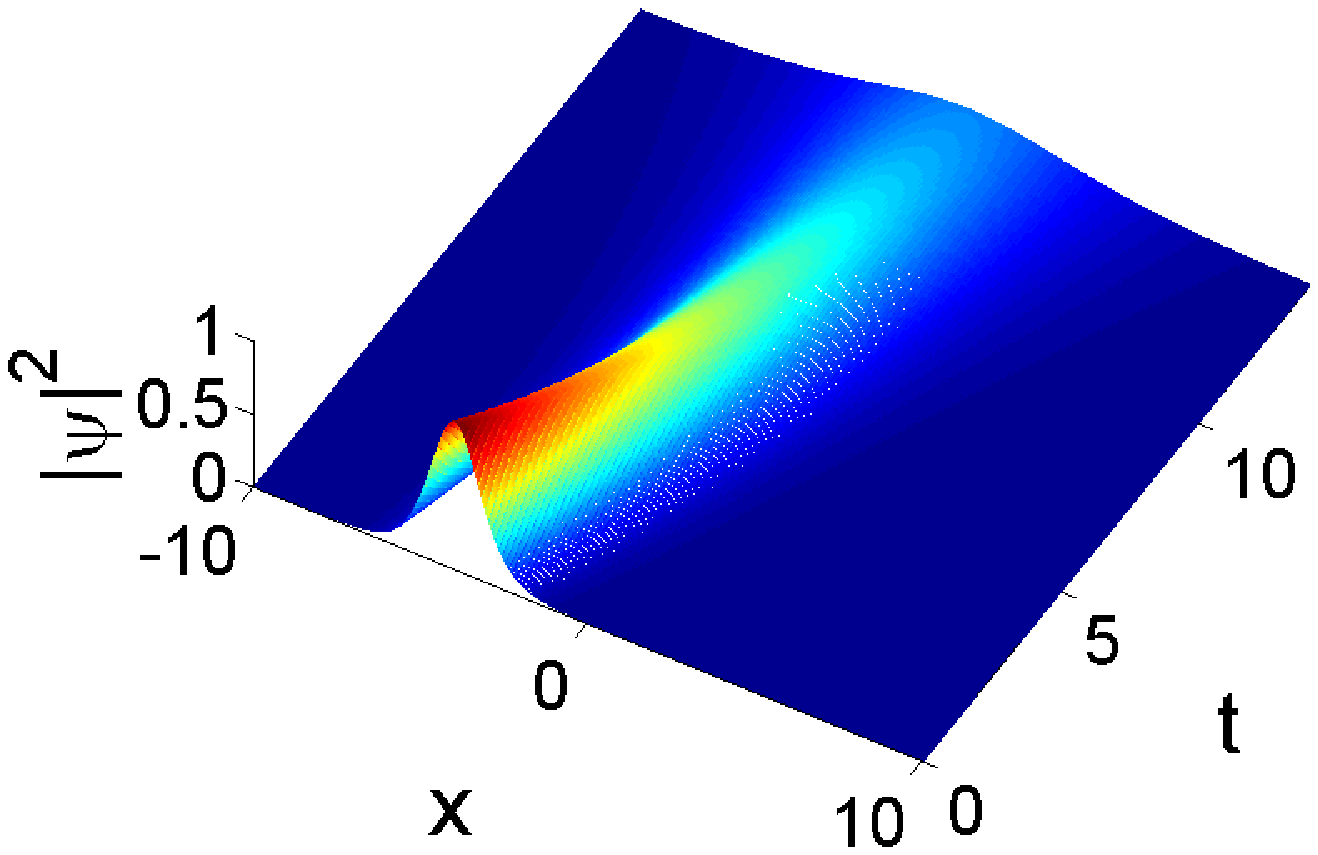}\textbf{(a)}
\includegraphics[width=2.0in]{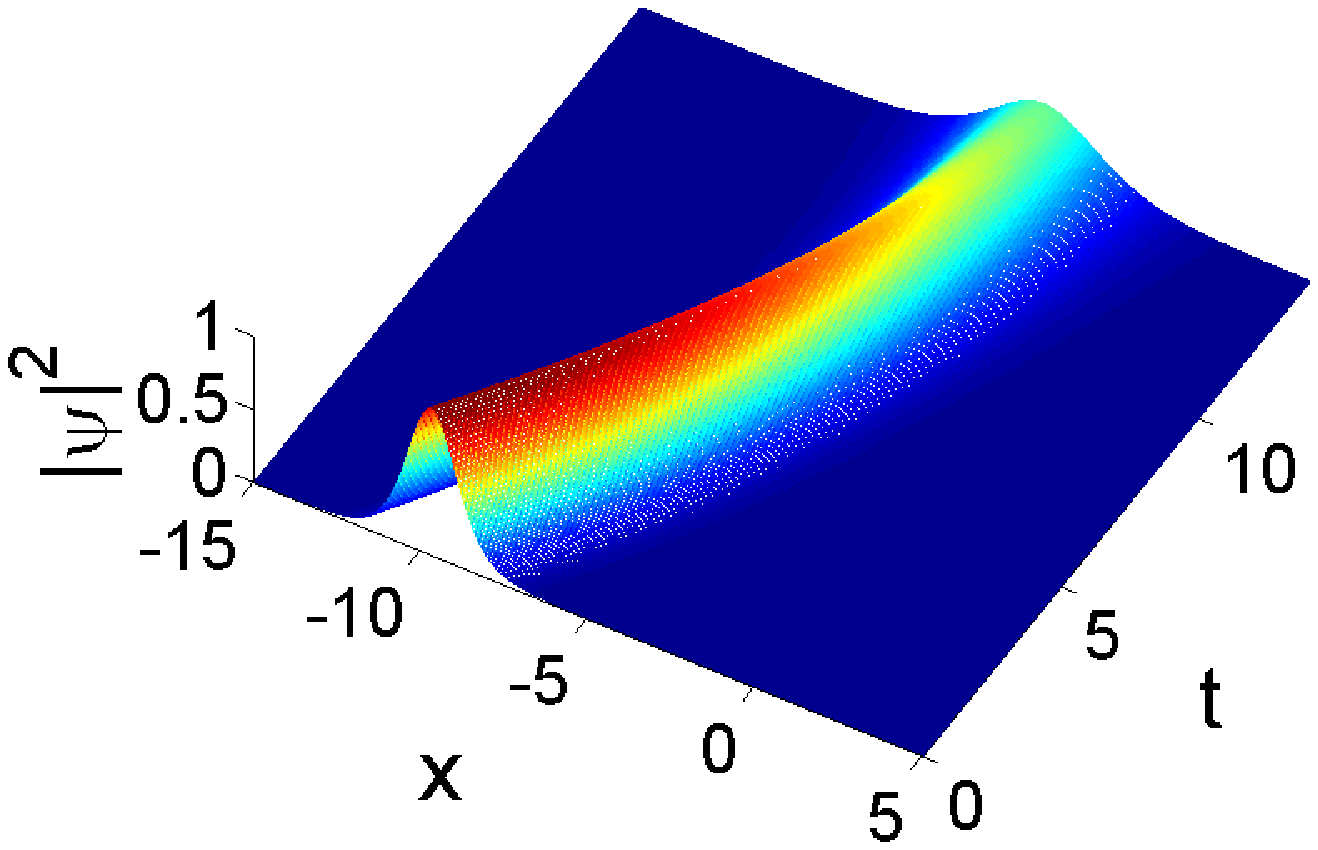}\textbf{(b)}
\includegraphics[width=2.0in]{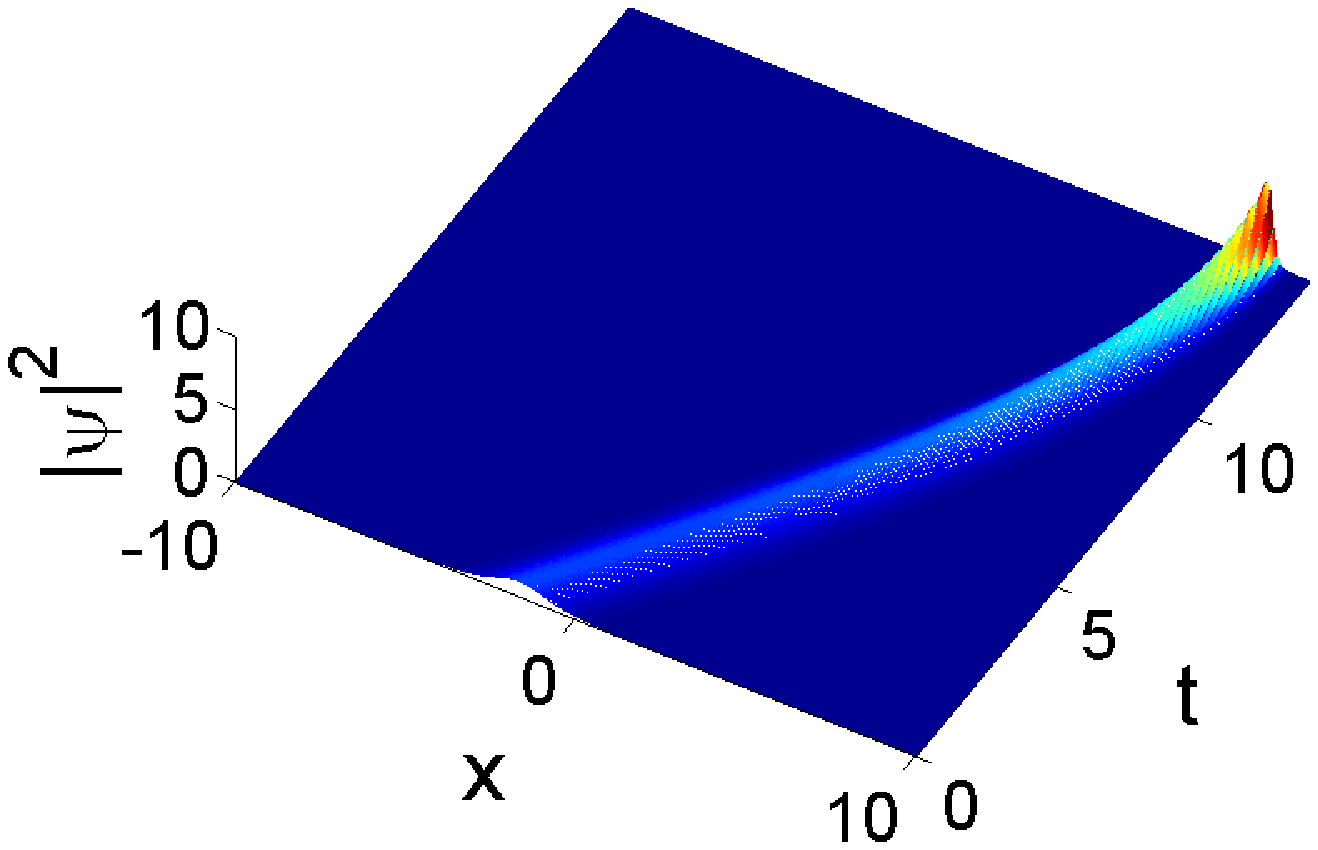}\textbf{(c)}
\caption{(Color online) Plot of the solution (\ref{eq38}) for (a)
$\beta(t)=\sqrt{-\frac{\alpha}{4c}}$, (b)
$\beta(t)=\sqrt{-\frac{\alpha}{4c}}\,\tanh(2\sqrt{-c\alpha}\,t)$,
and (c)
$\beta(t)=-\sqrt{\frac{\alpha}{4c}}\,\tan(2\sqrt{c\alpha}\,t)$. The
parameters are $\alpha=\pm0.005$, $p_{0}=1.0$, $\nu_{0}=1.0$, $a=0$,
$c=0.5$, $c_{2}=1$, $c_{4}=\nu_{0}$, $k_{1}=1.0$, $q_{1}=0$, and
$\lambda=0.01$. Then the cut-off time is approximately $t_{1}=15.7$.
All quantities are dimensionless.} \label{figbright}
\end{figure}

Considering the above results, the general solution of equation
(\ref{eq01}) for $c_{3}=0$ may be given by
\begin{equation}
\psi(z,t)=h(t)\,u(\zeta)\,e^{\mathrm{i}(\beta z^{2}+k z+\Omega)},
\label{eq38}
\end{equation}
where $u(\zeta)$ is the solution of Eq. (\ref{eq24}) given in Eq.
(\ref{eq25}). In order to obtain a soliton solution from Eqs.
(\ref{eq25}) and (\ref{eq38}), we need to avoid singularities in Eq.
(\ref{eq25}). Then, we assume in Eq. (\ref{eq25}) the constraint
$c_{4}<0$. In this case, taking $c_{3}=0$ as demanded above, one can
readily obtain
\begin{equation}
u(\zeta)=u_{0} \,\mathrm{sech}[\sqrt{c_{2}}\,(\zeta-\zeta_{0})],
\label{eq39a}
\end{equation}
where $u_{0}=\sqrt{-\frac{c_{2}}{c_{4}}}$ and
$\zeta_{0}=a-\frac{1}{2\sqrt{c_{2}}}\ln(-4c_{2}c_{4})$. In this case
Eq. (\ref{eq38}) describes a bright soliton. This solution is
displayed in Fig. \ref{figbright}. It should be noted that since
$c_{4}<0$, we have negative $\nu_{0}$ and then, this bright soliton
is meant for media with attractive interparticle interactions.

\subsection{The dark solitons}

Another case that may be of interest in obtaining soliton solutions
is when $c_{3}\neq 0$. We may have $f(t)\neq 0$, and reduce the set
of equations (\ref{eq29o}) to
\begin{equation}
\begin{split}
&h'+ 2c \beta h=0,\: p'+ 4c \beta p=0,\: q'+2c k p=0,\\
&k'+ 4c \beta k+\lambda=0,\: \beta'+4c\beta^{2}+\alpha=0,\\
&2cp^{2}c_{4}-\nu h^{2}=0,\: cp^{2}c_{3}-2 \nu f h=0,\: f'+ 2c \beta f=0,\\
&\Omega'+ck^{2}+\frac{1}{2}c p^{2}c_{2}=0,\:
 c p^{2}c_{2}-2\nu f^{2}=0. \label{eqa29}
\end{split}
\end{equation}

From Eq. (\ref{eqa29}), it may be shown that, when $c_{3}\neq 0$ the
parameters in Eq. (\ref{eq25}) must fulfill the constraint
$c_{3}^{2}=4c_{2}c_{4}$. In the case where $\alpha$ and $\lambda$
are constants, we choose the parameter $\beta(t)$ as in the previous
section to easily solve the above set of equations.

(1) When $\beta(t)$ is constant, the solution of Eq. (\ref{eqa29})
is:
\begin{equation}
\begin{split}
&\beta=\sqrt{\frac{-\alpha}{4c}},\, p(t)=p_{0} e^{-4c \beta t},\,
h(t)=h_{0} e^{-2c \beta t},\\
&f(t)=f_{0} e^{-2c \beta t},\, \nu(t)=\nu_{0} e^{-4c \beta t},\\
&k(t)=-\frac{\lambda}{4c \beta}+k_{1} e^{-4c \beta t},\\
&q(t)=\left(-\frac{\lambda}{8c\beta^{2}}+\frac{k_{1}}{4\beta}e^{-4c
\beta t}\right) p(t)+q_{1},\\
&\Omega(t)=-\frac{\lambda^{2}}{16c
\beta^{2}}t+\frac{c_{2}p_{0}^{2}+2k_{1}^{2}}{16\beta}e^{-8c \beta
t}-\frac{\lambda k_{1}}{8c\beta^{2}}e^{-4c \beta t}+\Omega_{1}.
\label{eqa30}
\end{split}
\end{equation}
In this set of equations, $f_{0}=\frac{c_{3}}{4c_{4}} h_{0}$ is the
initial value of $f(t)$. The difference between Eq. (\ref{eqa30})
and the corresponding results in the previous section resides in the
expressions for $f(t)$ and $\Omega(t)$ which change, respectively,
the amplitude and the phase of the matter wave.

(2) When
$\beta(t)=-\sqrt{\frac{\alpha}{4c}}\,\tan(2\sqrt{c\alpha}\,t)$,
solving the set of equations (\ref{eqa29}), we come to the following
results
\begin{equation}
\begin{split}
p(t)=& p_{0} \ell(t)^{-1},\, h(t)= h_{0} \ell(t)^{-1/2},\,
f(t)= f_{0} \ell(t)^{-1/2},\\
k(t)=& -\frac{\lambda}{2\sqrt{c\alpha}}\,\tan(2\sqrt{c\alpha}\,t)+k_{1} \ell(t)^{-1},\\
q(t)=& \left[-k_{1}\sqrt{\frac{c}{\alpha}}\sin(2\sqrt{c\alpha}\,t)+\frac{\lambda}{2\alpha}\right] p(t)+q_{1},\\
\Omega(t)=& \frac{\lambda^{2}}{4\alpha}t-\frac{\lambda^{2}+2c\alpha (c_{2} p_{0}^{2}+2k_{1}^{2})}{8\alpha\sqrt{c\alpha}}\,\tan(2\sqrt{c\alpha}\,t)\\
&+\frac{\lambda k_{1}}{2\alpha} \ell(t)^{-1}+\Omega_{1},\\
\nu(t)=& \nu_{0} \ell(t)^{-1},\, \ell(t)=
|\cos(2\sqrt{c\alpha}\,t)|. \label{eqa31}
\end{split}
\end{equation}
We can prevent singularities by considering only the times preceding
the cut-off time, i.e., we take $t<t_{1}$. However this particular
singularity can be avoided by changing the scattering length in a
small time interval that contains each singular time $t_{n}$.

\begin{figure}[tbh]
\centering
\includegraphics[width=2.0in]{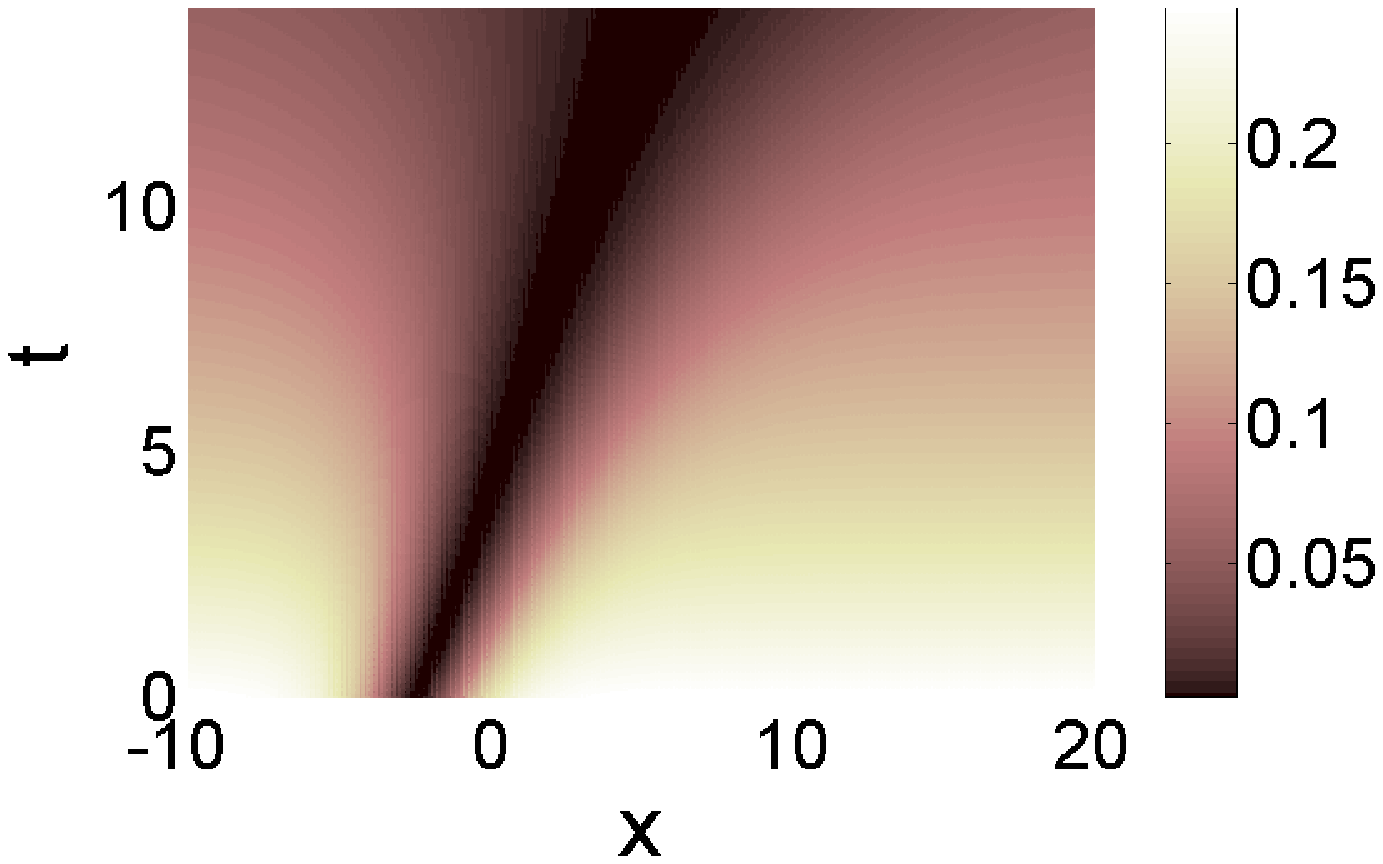}\textbf{(a)}
\includegraphics[width=2.0in]{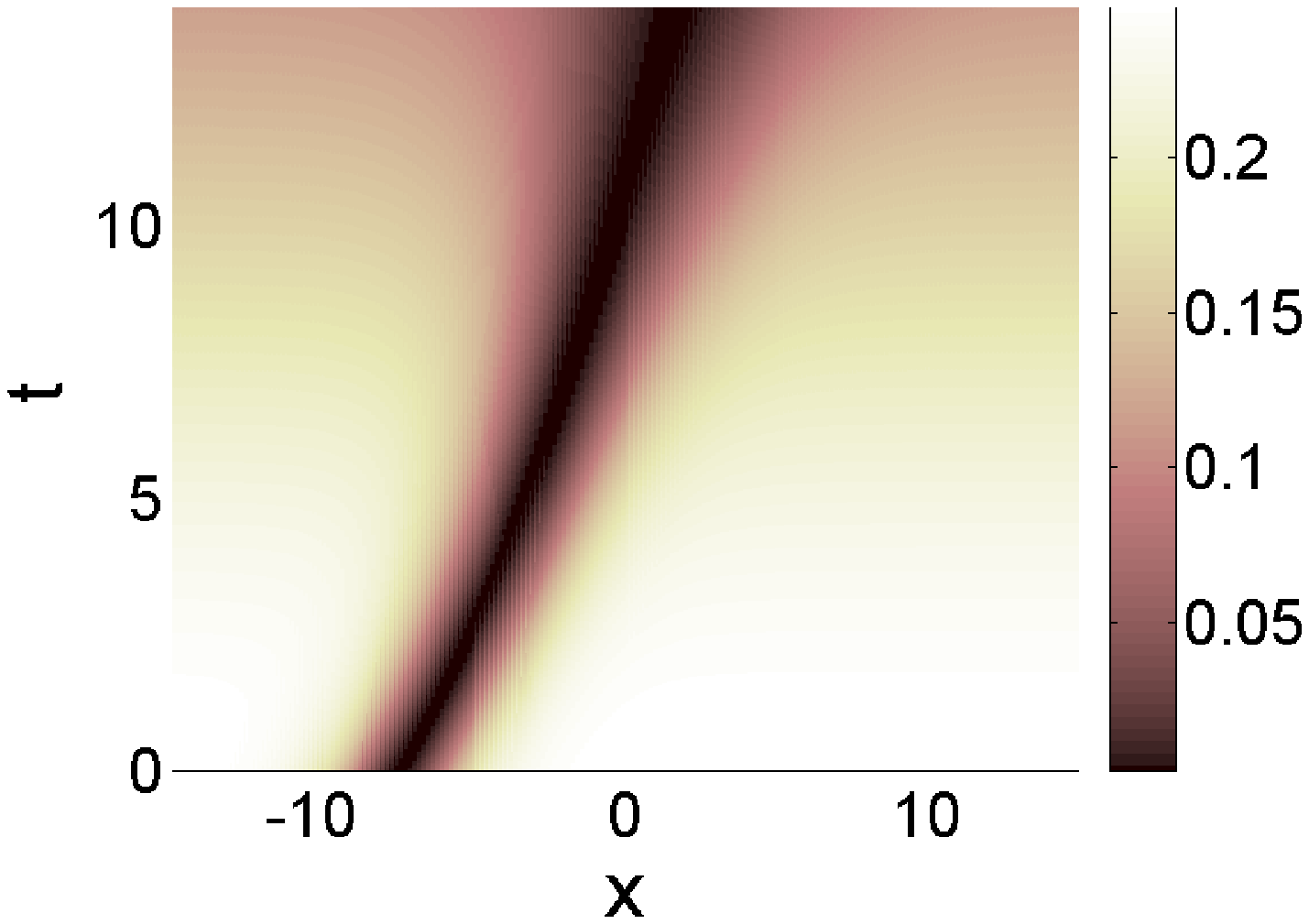}\textbf{(b)}
\includegraphics[width=2.0in]{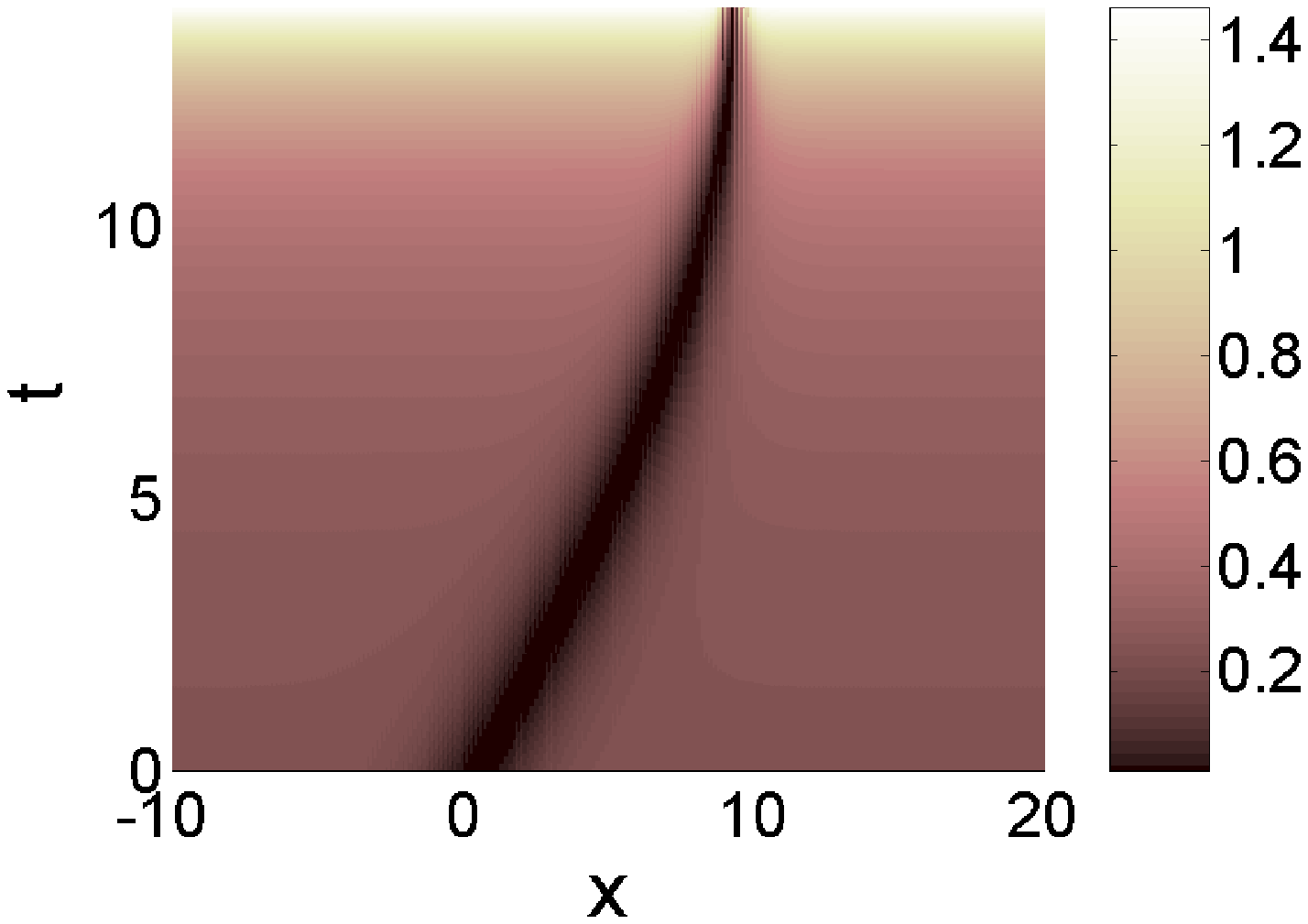}\textbf{(c)}
\caption{(Color online) Plot of the solution (\ref{eqa38}) for (a)
$\beta(t)=\sqrt{-\frac{\alpha}{4c}}$, (b)
$\beta(t)=\sqrt{-\frac{\alpha}{4c}}\,\tanh(2\sqrt{-c\alpha}\,t)$,
and (c)
$\beta(t)=-\sqrt{\frac{\alpha}{4c}}\,\tan(2\sqrt{c\alpha}\,t)$. The
parameters are $\alpha=\pm0.005$, $p_{0}=1.0$, $\nu_{0}=-1.0$,
$a=0$, $c=0.5$, $c_{2}=1$, $c_{4}=\nu_{0}$, $k_{1}=1.0$, $q_{1}=0$,
$\lambda=0.01$, and $\delta=1$. The cut-off time is roughly
$t_{1}=15.7$. All quantities are dimensionless.} \label{figdark}
\end{figure}

(3) When
$\beta(t)=\sqrt{-\frac{\alpha}{4c}}\,\tanh(2\sqrt{-c\alpha}\,t)$,
solving the set of equations (\ref{eqa29}), we obtain the following
results
\begin{equation}
\begin{split}
p(t)=& p_{0} \ell(t)^{-1},\, h(t)= h_{0} \ell(t)^{-1/2},\,
f(t)= f_{0} \ell(t)^{-1/2},\\
k(t)=& -\frac{\lambda}{2\sqrt{-c\alpha}}\,\tanh(2\sqrt{-c\alpha}\,t)+k_{1} \ell(t)^{-1},\\
q(t)=& \left(k_{1}\sqrt{-\frac{c}{\alpha}}\,e^{-2\sqrt{-c\alpha}\,t}+\frac{\lambda}{2\alpha}\right) p(t)+q_{1},\\
\Omega(t)=& \frac{\lambda^{2}}{4\alpha}t-\frac{\lambda^{2}+2c\alpha (c_{2} p_{0}^{2}+2k_{1}^{2})}{8\alpha\sqrt{-c\alpha}}\,\tanh(2\sqrt{-c\alpha}\,t)\\
&+\frac{\lambda k_{1}}{2\alpha} \ell(t)^{-1}+\Omega_{1},\\
\nu(t)=& \nu_{0} \ell(t)^{-1},\, \ell(t)=
\cosh(2\sqrt{-c\alpha}\,t). \label{eqa32}
\end{split}
\end{equation}
Considering the above results, the general solution of equation
(\ref{eq01}) for $c_{3}\neq 0$ is given by
\begin{equation}
\psi(z,t)=h(t)\left[\frac{c_{3}}{4c_{4}}+u(\zeta)\right]\,e^{\mathrm{i}(\beta
z^{2}+k z+\Omega)}, \label{eqa38}
\end{equation}
where $u(\zeta)$ is the solution of Eq. (\ref{eq24}) given in Eq.
(\ref{eq25}).
The condition $c_{3}^{2}=4c_{2}c_{4}$ implicitly assumes $c_{4}>0$
since $c_{2}$ is positive. In this case, we easily come to
\begin{equation}
u(\zeta)=\frac{4 c_{2}}{-2c_{3}+\:\exp[-\delta
\sqrt{c_{2}}\,(\zeta-a)]}. \label{eq39b}
\end{equation}
Eq. (\ref{eq39b}) is an interesting solution without singularity
only when the nonzero coefficient $c_{3}$ is negative. This
solution, with Eq. (\ref{eqa38}), corresponds to a darklike soliton.
We portray this solution in Fig. \ref{figdark}. We mention that
since $c_{4}>0$, the parameter $\nu_{0}$ is positive and
consequently this dark soliton is meant for media that present
repulsive interparticle interactions.

\section{The dynamical properties of the solutions} \label{sec4}

In order to investigate the dynamical properties of the solutions,
we reconsider the above three cases, namely the cases where
$\beta(t)=\sqrt{-\frac{\alpha}{4c}}$,
$\beta(t)=-\sqrt{\frac{\alpha}{4c}}\,\tan(2\sqrt{c\alpha}\,t)$, and
$\beta(t)=\sqrt{-\frac{\alpha}{4c}}\,\tanh(2\sqrt{-c\alpha}\,t)$
which correspond to the respective scattering lengths
$\upsilon(t)=\upsilon_{0} \exp(-2\sqrt{-c\alpha}\,t)$,
$\upsilon(t)=\upsilon_{0}/ |\cos(2\sqrt{c\alpha}\,t)|$, and
$\upsilon(t)=\upsilon_{0}/ \cosh(2\sqrt{-c\alpha}\,t)$.
The solution of equation (\ref{eq01}) in each of these cases, for
$c_{3}=0$ and $c_{3}\neq 0$, respectively, may read
\begin{equation}
\psi(z,t)=A_{0}
s(t)\,\mathrm{sech}[\sqrt{c_{2}}(\zeta-\zeta_{0})]\,e^{\mathrm{i}(\beta
z^{2}+k z+\Omega)}, \label{eq38a}
\end{equation}
\begin{equation}
\psi(z,t)=A_{0}
s(t)\tanh[\frac{1}{2}\sqrt{c_{2}}(\zeta-\zeta_{1})]e^{\mathrm{i}(\beta
z^{2}+k z+\Omega)}. \label{eqb38a}
\end{equation}
The time-dependent amplitude coefficient, $s(t)=\ell(t)^{-1/2}$,
depends on the typical forms of $\beta(t)$ into consideration. We
have $A_{0}=\sqrt{\frac{2c c_{2}}{|\nu_{0}|}p_{0}^{2}}$ and
$\zeta_{1}=\zeta_{0}-\frac{\ln 2}{\sqrt{c_{2}}}$. Additionally, we
recall that $\zeta=p\,z+q$ and
$\zeta_{0}=a-\frac{1}{2\sqrt{c_{2}}}\ln(4c_{2}|c_{4}|)$. Without
loss of generality, we set $q_{1}=0$.

\subsection{The vanishing matter waves}

When $\beta=\sqrt{-\frac{\alpha}{4c}}$, the solution of equation
(\ref{eq01}) is given in Eqs. (\ref{eq38a}) and (\ref{eqb38a}),
where $\ell$, $p$, $q$, $\beta$, $k$ and $\Omega$ are time-dependent
functions given in Eq. (\ref{eq30}) for the solution in Eq.
(\ref{eq38a}), and in Eq. (\ref{eqa30}) for the solution in Eq.
(\ref{eqb38a}). As already said, these solutions represent bright
and dark solitons, respectively. However they have common dynamical
behavior. The width of each of them is proportional to $(p_{0}^{2}
c_{2})^{-1/2}\,e^{4c\beta t}$, while the height is proportional to
$A=A_{0}\,e^{-2c\beta t}$. In this case, the matter waves have
broadening and vanishing properties. As a matter of fact, with time,
the height of the matter wave decreases while its width increases.
However the number of atoms in the condensate, i.e.
$N=\int|\psi|^{2}\mathrm{d}z=-\frac{4c\sqrt{p_{0}^{2}
c_{2}}}{\nu_{0}}$, remains unchanged during the propagation of the
wave. A plot of this dynamical behavior is given in Figs.
\ref{figbright}(a) and \ref{figdark}(a) through the space-time
evolution of the square magnitude of the wave function.

The kinematics of the wave can be obtained from $\zeta(z,t)=0$.
Hence the motion of the center of mass, taken as the position that
corresponds to the peak, is determined by the following equation:
\begin{equation}
z=-\frac{1}{2}
k_{1}\sqrt{-\frac{c}{\alpha}}\,e^{-2\sqrt{-c\alpha}\,t}-\frac{\lambda}{2\alpha},
\label{eqb38b}
\end{equation}
for both dark and bright solitons. At longer times, the center of
mass of the wave is driven towards the point
$z=-\frac{\lambda}{2\alpha}\equiv z_{\infty}$ which corresponds to
the effective trap center. In fact, when the gravitational field is
considered, the minimum of the potential is no more on the magnetic
trap axis $z=0$, it moves to $z_{\infty}$. Hence the gravitational
field drives the wave from the center of the magnetic trap to a
region around $z_{\infty}$, where the wave should be confined.

The velocity of the wave packet is $\dot{z}=c
k_{1}\,e^{-2\sqrt{-c\alpha}\,t}$. Hence $k_{1}$ appears to be a
measurement of the initial velocity of the wave. The velocity
exponentially decreases with time. So the choice of the parameters
$\lambda$ and $k_{1}$ can seriously affect the dynamics of the
matter waves, denouncing the role of the gravitational field in our
analysis. The wave packet behaves like a static classical particle
for $k_{1}=0$ and like a moving one for $k_{1}\neq 0$. A similar
result was obtained in \cite{li1} within a special case where the
gravitational field is absent ($\lambda=0$) and $c_{3}= 0$. The
acceleration of the wave packet is $\ddot{z}=-2\sqrt{-c^{3}\alpha}
k_{1}\,e^{-2\sqrt{-c\alpha}\,t}$. The acceleration exponentially
decreases with time.

When $\beta=\sqrt{-\frac{\alpha}{4c}}\,\tanh(2\sqrt{-c\alpha}\,t)$,
the solution of equation (\ref{eq01})
is given in Eqs. (\ref{eq38a}) and (\ref{eqb38a}), where $\ell$,
$p$, $q$, $\beta$, $k$ and $\Omega$ are time-dependent functions
given in equation (\ref{eq32}) for the solution in Eq.
(\ref{eq38a}), and in equation (\ref{eqa32}) for the solution in Eq.
(\ref{eqb38a}).
These solutions also represent growing matter waves. The height of
each of these waves is proportional to $\left(\frac{2c
p_{0}^{2}c_{4}}{\nu_{0}\cosh(2\sqrt{c\alpha}\,t)}\right)^{1/2}$, and
the width is proportional to
$\frac{1}{p_{0}}\cosh(2\sqrt{c\alpha}\,t)$. The width of the soliton
shortens exponentially with time while its height exponentially
grows. A display of this dynamical behavior can be found in Figs.
\ref{figbright}(b) and \ref{figdark}(b) where we plot the space-time
evolution of the wave in the system.
The motion of the center of mass of the matter wave is defined by
the equation:
\begin{equation}
z=-k_{1}\sqrt{-\frac{c}{\alpha}}\,e^{-2\sqrt{-c\alpha}\,t}-\frac{\lambda}{2\alpha}.
\label{eqb38c}
\end{equation}
At longer times, the center of mass of the wave is driven towards
the point $z=\frac{\lambda}{2\alpha}\equiv z_{\infty}$ which
corresponds to the effective trap center. The velocity of the wave
packet is $\dot{z}=2c k_{1}\,e^{-2\sqrt{-c\alpha}\,t}$, which is
twice the velocity in the previous case.
We have observed that the solutions in the cases
$\beta(t)=\sqrt{-\frac{\alpha}{4c}}$ and
$\beta(t)=\sqrt{-\frac{\alpha}{4c}}\,\tanh(2\sqrt{-c\alpha}\,t)$,
both corresponding to an expulsive trapping potential, present
similar asymptotic behavior in time. In fact, when $t \rightarrow
\infty$ the width, height, and trajectory in both cases become
identical.

\subsection{The growing matter waves}

\begin{figure}[tbh]
\centering
\includegraphics[width=2.0in]{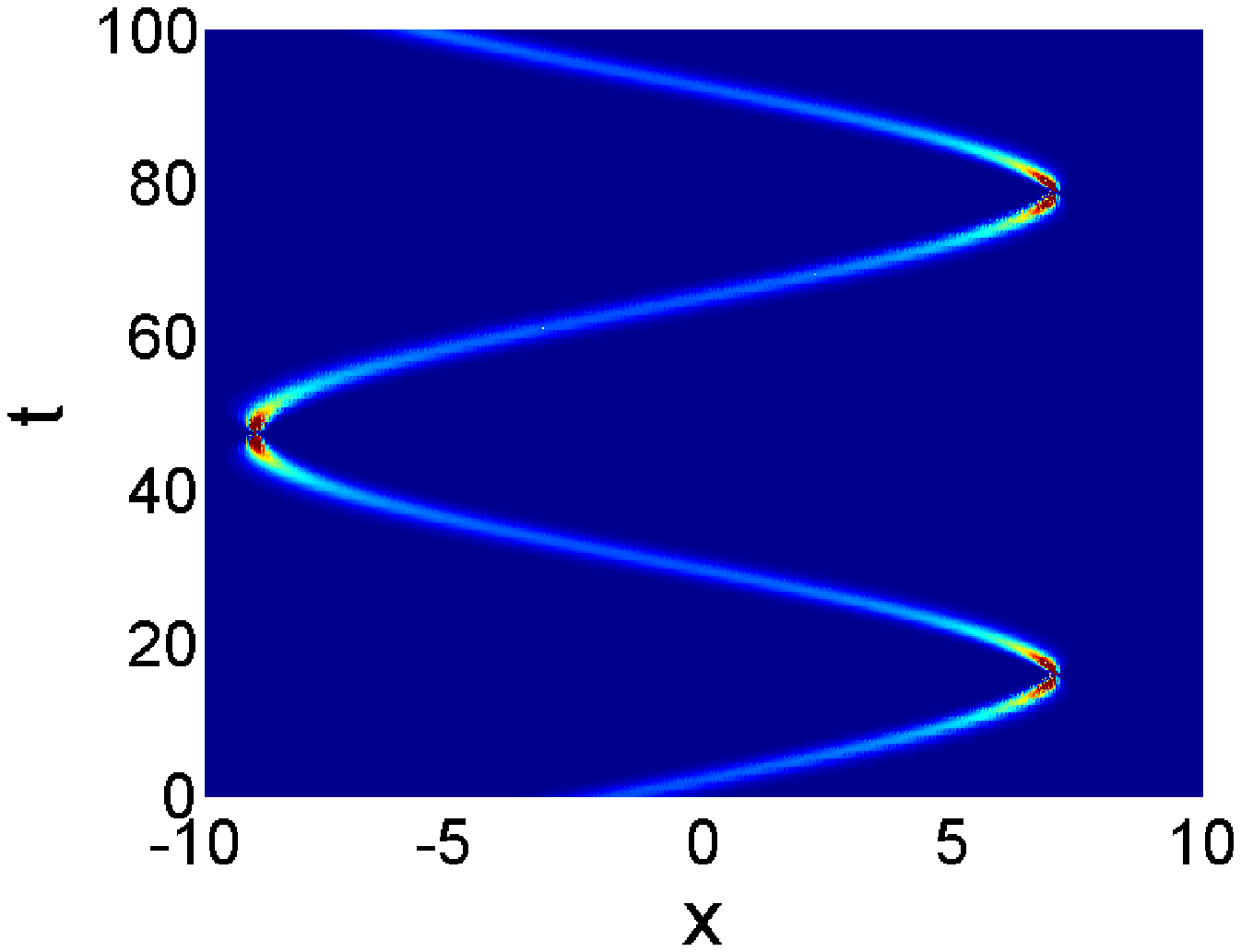}\textbf{(a)}
\includegraphics[width=2.0in]{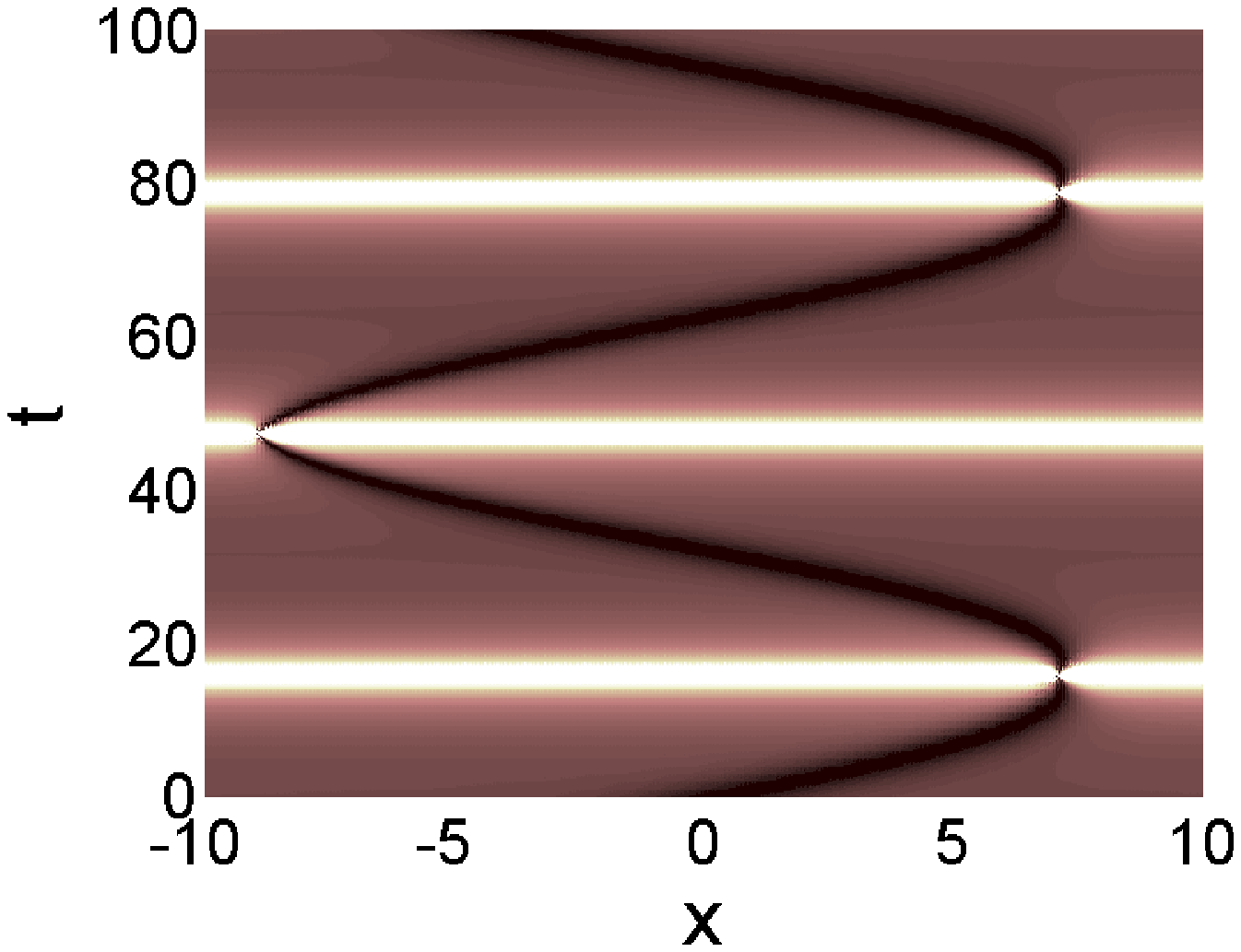}\textbf{(b)}
\caption{(Color online) Plot of the long-time evolution of the
matter waves in the case where
$\beta(t)=-\sqrt{\frac{\alpha}{4c}}\,\tan(2\sqrt{c\alpha}\,t)$ for
(a) the solution in Eq. (\ref{eq38}), and (b) the solution in Eq.
(\ref{eqa38}). We use
$\beta(t)=\sqrt{-\frac{\alpha}{4c}}\,\tanh(2\sqrt{-c\alpha}\,t)$ in
small time intervals that contain each singular time
$t_{n}=\frac{(2n+1)\pi}{4\sqrt{c\alpha}}$, with $n$ being any
integer. The parameters are $\alpha=\pm0.005$, $p_{0}=1.0$,
$\nu_{0}=-1.0$, $a=0$, $c=0.5$, $c_{2}=1$, $c_{4}=\nu_{0}$,
$k_{1}=0.8$, $q_{1}=0$, $\lambda=0.01$, and $\delta=1$. All
quantities are dimensionless.} \label{figlong}
\end{figure}

When $\beta=-\sqrt{\frac{\alpha}{4c}}\,\tan(2\sqrt{c\alpha}\,t)$,
the solution of equation (\ref{eq01})
is given in Eqs. (\ref{eq38a}) and (\ref{eqb38a}), where $\ell$,
$p$, $q$, $\beta$, $k$ and $\Omega$ are time-dependent functions
given in equation (\ref{eq31}) for the solution in Eq.
(\ref{eq38a}), and in equation (\ref{eqa31}) for the solution in Eq.
(\ref{eqb38a}). The corresponding solutions represent growing
 matter waves. The height of each of these matter waves is proportional to
$\left(\frac{2c
p_{0}^{2}c_{4}}{\nu_{0}|\cos(2\sqrt{c\alpha}\,t)|}\right)^{1/2}$,
and the width is proportional to
$\frac{1}{p_{0}}|\cos(2\sqrt{c\alpha}\,t)|$.  In the safe time
interval, the matter wave becomes thinner and higher.  We portray in
Figs. \ref{figbright}(c) and \ref{figdark}(c) this dynamical
behavior. Close to the cut-off time which is $t_{1}=15.7$, the
exponential increase in the amplitude of the wave is so strong that
a "collapse" of the wave may occur. However, by changing the
expression of the parameter $\beta(t)$ (which amounts to changing
the expression of the \emph{s}-wave scattering length) in a small
time interval that contains each singular time
$t_{n}=\frac{(2n+1)\pi}{4\sqrt{c\alpha}}$, the propagation of the
matter wave can be kept. It can be changed to
$\beta(t)=\sqrt{-\frac{\alpha}{4c}}$ or
$\beta(t)=\sqrt{-\frac{\alpha}{4c}}\,\tanh(2\sqrt{-c\alpha}\,t)$. In
this case, the width and the height of the wave oscillate in time.
Figures \ref{figlong}(a) and \ref{figlong}(b) show the long-time
propagation of the matter waves in this case.

The motion of the center of mass of the matter wave is defined by
the equation:
\begin{equation}
z=k_{1}\sqrt{\frac{c}{\alpha}}\sin(2\sqrt{c\alpha}\,t)-\frac{\lambda}{2\alpha}.
\label{eqb38d}
\end{equation}
The velocity of the wave packet is $\dot{z}=2c
k_{1}\cos(2\sqrt{c\alpha}\,t)$ while its acceleration is
$\ddot{z}=4\sqrt{c^{3}\alpha} k_{1}\cos(2\sqrt{c\alpha}\,t)$. This
means that the wave oscillates in time with frequency
$f=\frac{\sqrt{c\alpha}}{\pi}$ also equivalent to
$(t_{n+1}-t_{n-1})^{-1}$. These oscillations are performed around
the position $z=-\frac{\lambda}{2\alpha}$, which corresponds to the
effective trap center set by the gravitational field.

In comparison with the results obtained in \cite{wamba8}, we see
that the role of gravity would not be the same, even qualitatively,
for different traps of the BEC system. In fact, the bias magnetic
field may be the analog of gravitational field since both fields are
represented by the linear term in the trapping potential. From the
results of \cite{wamba8}, we infer that the kink solitons created in
a bias magnetic field alone behave like a classical particle in a
pure free fall motion led by the "gravity" in the $(z,t)$ space. In
the present case, the gravitational field is not alone. The presence
of the parabolic magnetic potential changes the effect of the
gravitational field. For instance, when the initial speed of the
bright or dark soliton is zero, then both the velocity and the
acceleration at any time are zero too.

\par

The present study suggests three ways to generate bright and dark
solitons in BEC systems by time-varying the \emph{s}-wave scattering
length (through the Feshbach resonance) without changing the
magnetic potential. This can be done by tuning the scattering length
to $g_{0}/ |\cos(2\sqrt{c\alpha}\,t)|$ when the condensate is
confined in an attractive magnetic trap, i.e. $\alpha$ is positive.
We may also tune the scattering length to $g_{0}/
\cosh(2\sqrt{-c\alpha}\,t)$, or $ g_{0} \exp(-2\sqrt{-c\alpha}\,t)$,
in the case of expulsive magnetic trap, i.e. $\alpha$ is negative.

\section{Conclusion}\label{sec5}

\ In conclusion, we have considered the GP equation with
time-dependent cubic nonlinearity which describes the dynamics of
the BEC matter-waves in a magnetic field and under the effect of a
homogeneous gravitational field. With the help of the extended
tanh-function method, we have obtained a solution which has as
special cases the bright and dark solitons. As has been discussed,
these solitons can be generated by properly tuning the \emph{s}-wave
scattering length of the condensed particles, depending on whether
the magnetic trapping is attractive or expulsive. The dynamics and
kinematics of these matter waves have been presented and discussed.

We have found that the gravity reshapes the repel force of the
magnetic trap, and then drives the matter waves towards the region
around the position $z=-\frac{\lambda}{2\alpha}$. The matter waves
may remain in that position for scattering lengths $g_{0}/
\cosh(2\sqrt{-c\alpha}\,t)$ or $ g_{0} \exp(-2\sqrt{-c\alpha}\,t)$,
and may oscillate around it for $g_{0}/ |\cos(2\sqrt{c\alpha}\,t)|$.
By comparing the results obtained here with those of \cite{wamba8},
we have found that the role of gravitational field depends on the
type of trap in which the condensate is confined.

The study of the dynamics and stability of a BEC under the effect of
very strong gravitational field, that could occur (in a speculative
way) for instance close to black holes, appears to pose an
interesting issue to investigate in future works.

\section*{Acknowledgments}

Part of this work has been done during the Short Term Visit of EW
within the CMSP Section of the Abdus Salam ICTP (Italy). EW
acknowledges the support from the Government of India, through the
CV Raman International Fellowship for African Researchers. HV
acknowledges the support from CNPq (Brazil) and the DFG Research
Training group 1620 "Models of Gravity". AM Thanks the Abdus Salam
ICTP for financial support through the Associateship program. KP
acknowledges DST, DAE-BRNS, UGC, the Government of India, for
financial support through major projects.

\end{document}